\documentclass[%
 reprint,
 amsmath,amssymb,
 aps,
]{revtex4-2}

\usepackage{graphicx}
\usepackage{dcolumn}
\usepackage{bm}

\usepackage{color}
\usepackage{pgfplots,mathtools}
\usepackage{hyperref}

\usepackage{braket}
\usepackage{slashed}
\usepackage[compat=1.0.0]{tikz-feynman}
\newcommand{\be}{\begin{equation}}
\newcommand{\ee}{\end{equation}}
\newcommand{\bea}{\begin{eqnarray}}
\newcommand{\eea}{\end{eqnarray}}
\newcommand{\ba}[1]{\begin{array}{#1}}
\newcommand{\ea}{\end{array}}
\newcommand{\nn}{\nonumber}

\newcommand{\ep}{\epsilon}

\newcommand{\del}{\partial}

\begin{document}
\title{Shear viscosity expression for a graphene system in relaxation time approximation}
\author{Cho Win Aung, Thandar Zaw Win, Gaurav Khandal, Sabyasachi Ghosh}
 \affiliation{Indian Institute of Technology Bhilai, GEC Campus, Sejbahar, Raipur 492015, Chhattisgarh, India}
\begin{abstract}
We have gone through the detailed microscopic calculation of the shear viscosity of a 2-dimensional graphene system in the 
relaxation time approximation-based kinetic theory framework. After getting its final expressions, we compared
it with the shear viscosity expressions of other possible 2-dimensional as well as 3-dimensional nonrelativistic and 
ultra-relativistic fluid systems. The aim of the comparison is to reveal - how their different one-body dispersion relations
affect their many-body fluid properties like shear viscosity and the viscosity to entropy density ratio. It is also aimed to
reveal the 3-dimension to the 2-dimension transformation of their mathematical structures. We have numerically
explored the differences in their order of magnitude and dependence on thermodynamical parameters \textendash{} temperature and chemical
potential. Marking two thermodynamical domains \textendash{} Dirac fluid and Fermi liquid \textendash{} for a 2-dimensional graphene system, we have noticed
that shear viscosity, entropy density as well as their ratios decrease toward saturated values when one goes from Fermi liquid
to Dirac fluid domain. When one shifts from mili-electron volt scales of temperature and chemical potential in condensed matter physics
location to their Mega-electron volt scales in high energy physics location, then the same results may be expected for hot quark matter case,
where the transition from the neutron star to early universe domains may be considered as Fermi liquid to Dirac fluid transition.

\end{abstract}
\maketitle
\section{Introduction} 
    It is known that the mean free path of charge carriers in metal is generally temperature dependent. The scattering between electron and lattice imperfections (or “disorder”) normally dominates at low temperatures, while electron-phonon scatterings dominate at high temperatures. Concerning these two scattering mechanisms, another possible scattering is electron-electron scattering processes, which are generally less effective in many conventional metals. However, its opposite condition is possible in some specific systems under specific conditions, where one can apply the electron hydrodynamic (eHD) theory. For a long time, condensed matter physicists did not aware of such an opposite phase in materials. Therefore, they used to give less attention to the possibilities of the hydrodynamics behavior of electrons. After the experimental observations of eHD in Refs.~\cite{1Ku:2019lgj,2varnavides2020electron,3PhysRevB.103.125106,4PhysRevB.103.235152,5PhysRevB.103.155128,6PhysRevResearch.3.013290,7DiSante:2019zrd,8PhysRevB.103.115402,9sulpizio2019visualizing,10gallagher2019quantum,11doi:10.1126/science.aau0685,12ella2019simultaneous,13bandurin2018fluidity,18doi:10.1126/science.aad0201,14PhysRevB.98.241304,15jaoui2018departure,16doi:10.1126/science.aac8385,17doi:10.1126/science.aad0343}~, the situation has drastically changed in recent years. See Refs.~\cite{eHD1,eHD2,eHD3} for recent reviews. It is graphene \cite{1Ku:2019lgj,2varnavides2020electron,3PhysRevB.103.125106,4PhysRevB.103.235152,5PhysRevB.103.155128,6PhysRevResearch.3.013290,7DiSante:2019zrd,8PhysRevB.103.115402,9sulpizio2019visualizing,10gallagher2019quantum,11doi:10.1126/science.aau0685,12ella2019simultaneous,13bandurin2018fluidity,18doi:10.1126/science.aad0201}, which is identified as the best known such material, where electron hydrodynamics can be observed. Apart from these recently discovered hydrodynamic properties of electrons in graphene, it was quite famous for its massless nature, concluded from the proportional relation between its energy and momentum. Due to the proportional relation between energy and momentum, electron motion in graphene will not be Galilean-invariant. On the other hand, the relativistic effect of electrons can not also be expected because its velocity ($v_g\approx 10^6$ m/s) is not very close to the speed of light ($c\approx 3\times 10^8$ m/s). Hence, we cannot claim the Lorentz-invariant property of electron motion. It opens “unconventional” hydrodynamics~\cite{eHD2} as neither nonrelativistic hydrodynamics (NRHD) nor relativistic hydrodynamics (RHD) can be applicable. We may call this “unconventional” hydrodynamics as Graphene hydrodynamics (GHD) by imposing that the graphene (G) case has a unique dispersion or energy-momentum relation and is different from the nonrelativistic (NR) and relativistic (R) or ultra-relativistic (UR) cases.
    Now, whenever fluid dynamics or hydrodynamics comes into the picture, then one dissipation coefficient like the shear viscosity of that fluid, becomes a very important quantity, which is not at all appeared in most of the metals or other condensed matter systems. The present work is aimed at the microscopic calculation of the shear viscosity of this electron fluid in a graphene system, which may be called in short as graphene fluid (GF). When one microscopically calculates the expression of the shear viscosity of GF, it will be different from its standard expression for nonrelativistic fluid (NRF) as well as for relativistic fluid (RF) or ultra-relativistic fluid (URF).    
    So far, from the best of our knowledge, experimental measurement of this shear viscosity coefficient for GF is missing although experimental community~\cite{1Ku:2019lgj,9sulpizio2019visualizing} observed the Poiseuille's flow pattern of electrons in graphene, which indirectly reflects the existence of the non-zero viscosity. From the theoretical side, we get only Refs.~\cite{24PhysRevLett.103.025301, Vignale, PhysRevB.97.Russian, PhysRevB.107.Ele_Vis, PhysRevB.100.Magneto}, where microscopic expressions of shear viscosity have been addressed. In this context, one can get a long list of Refs.~\cite{QGP1, QGP2, QGP3, QGP4, QGP5, QGP6, QGP7, QGP8, QGP9, QGP10} (and references therein) for microscopical estimations of shear viscosity for relativistic quark and hadronic matter, expected in high energy heavy ion collision experiments. Grossly two classes of frameworks - (1) Kinetic theory approach with relaxation time approximation (RTA)~\cite{QGP1, QGP2, QGP3, QGP4, QGP5, QGP6, QGP7} and Kubo framework~\cite{QGP8, QGP9, QGP10} - are adopted by the heavy ion physics community. Both frameworks have similar structure at the final level expressions for shear viscosity coefficients with two main components. One carries interaction information, called relaxation time, and the remaining part may be called as thermodynamic phase-space of shear viscosity coefficient, which will be function of temperature and chemical potential. If we analyze the shear viscosity expression of graphene also from Ref.~\cite{24PhysRevLett.103.025301}, then we can identify these two components. Present work has zoomed in this structure via a systematic calculation of this shear viscosity of GF in RTA methods and compared with corresponding structures for NRF and URF. Here, one of our aims is to compare the thermodynamic phase-space component of shear viscosity coefficient for these three cases - G, NR, and UR.  
    After knowing the lower bound conjecture of shear viscosity to entropy density $(\eta/s)$ as $\hbar/(4\pi k_B)$ or $1/(4\pi)$ (in natural unit)~\cite{25Kovtun:2004de}, scientific communities are curious to know those strongly coupled systems, which are close to that bounds. Experimentally, the RF, like quark and hadronic matter, produced in high energy heavy ion collision experiments and NRF like cold-atom systems~\cite{27Sch_fer_2009} are identified as those strongly coupled systems. On the other hand, GF may also belong to that category according to the theoretical prediction from Ref.~\cite{24PhysRevLett.103.025301}, which is considered as reference point for tuning our results. So, the present article will not intend to add any new content on strongly coupled properties, rather its main goal will intend to find the differences among GF, NRF, URF in terms of expressions and estimations of shear viscosity. We believe that it was missing in the literature and very important to address.
%
    The article is organized as follows. In the next section Sec.~(\ref{sec: Form}), the RTA calculation of shear viscosity $\eta$ and entropy density $s$ calculations of GF for 2D case is addressed in detail by mentioning the other cases like 3D-GF, 3D-NR, 3D-UR, 2D-NR, 2D-UR. In Sec.~(\ref{sec: Result}), the comparative results of $\eta$, $s$ and $\eta/s$ of different cases are discussed. At the end, our findings are summarised in Sec.~(\ref{sec: Sum}) with some conclusive bullet points.
%
    \section{Formalism}
    \label{sec: Form}
    Let us start our formalism from energy-momentum tensor ($T^{\mu\nu}$), as practised for RF like quark and hadronic matter. Here, we will go for GF calculation, so reader should have to be careful on some particular steps, where it is different from the RF case. Showing these differences is one of the core agenda of the present article. Although, reader can find similarities between most of steps of GHD of GF and RHD of RF. The $T^{\mu\nu}$ has two parts - the ideal part $T_0^{\mu\nu}$, related to the knowledge of thermodynamics and the dissipative part $T_D^{\mu\nu}$, related to the different dissipation processes. So,
    \bea
    T^{\mu\nu} &=& T_0^{\mu\nu} + T_D^{\mu\nu}. 
    \eea
    In this dynamic picture of fluid, ideal energy-momentum tensor and electron number flow can be expressed in macroscopic form as,
    \bea
    T_0^{\mu\nu} &=&  \ep \frac{u^\mu u^\nu}{v_g^2} - P\left(g^{\mu\nu} -\frac{u^\mu u^\nu}{v_g^2}\right),
    \nn\\
    N_0^\mu &=& n \frac{u^\mu}{v_g},
    \label{eq2}
    \eea
    in terms of the building blocks - energy density $\epsilon$, pressure $P$, number density $n$, fluid (element) velocity $u^\mu$ and metric tensor $g^{\mu \nu}$. Here, four-velocity $u^{\mu}=\gamma_g (v_g, \vec{u})$ for GHD is designed by following the four velocity structure $u^{\mu}=\gamma (c, \vec{u})$ for RHD as done in Ref.~\cite{eHD2}. One can notice that speed of light $c$ in RHD is replaced by graphene Fermi velocity $v_g$ in GHD. So, Lorentz factor $\gamma= 1/\sqrt{1-u^2/c^2}$ in RHD is also converted into $\gamma_g= 1/\sqrt{1-u^2/v_g^2}$ in GHD. In static limit (${\vec u}\rightarrow 0$), four velocity, $u^{\mu}=\gamma_g (v_g, \vec{u}) \rightarrow u^{\mu}=\gamma_g (v_g, 0)$ and $\gamma_g= 1/\sqrt{1-u^2/v_g^2} \rightarrow 1$. So, Eq.~(\ref{eq2}) provides a static electron number flow $N^\mu_0\equiv n$, and static energy-momentum tensor, 
    \begin{equation}
    T_0^{\mu\nu} \equiv  \begin{pmatrix}
    \epsilon &  0   &  0 & 0\\
     0  &   P &  0 & 0\\
     0  &  0  &  P &  0\\
     0  &  0  &  0 &  P\\
     \end{pmatrix},
    \end{equation}
    which reflects  the standard static fluid aspect like Pascal's law. The macroscopic quantities $T_0^{\mu \nu}$ and $N^\mu_0$ can be expressed in terms of the microscopic quantities; four-momentum ($p^{\mu}$) and four-velocity ($v^{\mu}$) of electrons as,
    \begin{equation}
    T_0^{\mu\nu} = N_s\int \frac{d^3\vec{p}}{(2\pi)^3}p^\mu v^\nu f_0,
    \label{rohan}
    \end{equation}
    and
    \begin{equation}
    N_0^{\mu} = N_s\int \frac{d^3\vec{p}}{(2\pi)^3} v^\mu f_0,
    \label{N_mic}
    \end{equation}
    where $N_s=2$ is spin degeneracy factor of electron and $f_0$ is its Fermi-Dirac (FD) distribution function $f_0=1/\{\exp{(\beta(E-\mu))}+1\}$. Here, $\beta=1/(k_B T)$ and $\mu$ are the thermodynamic parameter and the chemical potential of the system, respectively. From these microscopic expressions of the energy-momentum tensor and electron current, given in Eqs.~(\ref{rohan}) and (\ref{N_mic}), we can write the energy density $\ep$, pressure $P$ and number density $n$ for 2D graphene (G) case, which is addressed briefly in next subsection. We follow natural unit $\hbar=c=k_B=1$ during the calculation.
    \subsection{Entropy density in two-dimensional Graphene}

    For Graphene, the dispersion relation is given by
    \begin{equation}
    E = pv_g~.
    \label{nanu}
    \end{equation}
    The total number of fermions at any value of temperature is given by,
    \begin{equation}
    N = \int_0^{\infty}D\left(E\right)dE f_0,
    \label{3dnd}
    \end{equation}
    where $D\left(E\right)dE$ is number of energy states in energy range $E$ to $E+dE$.
    After plugging the value of $D\left(E\right)dE$ (see Appendix~\ref{App_A}) in the above Eq.~(\ref{3dnd}) and $f_0$, the total number of electrons in graphene is,
    \begin{align*}
    &N = N_s \frac{2\pi a}{\left(2\pi\right)^2 v_g^2} \int_{0}^{\infty}  \frac{E}{A^{-1}e^{\beta E} + 1} dE,
    \end{align*}
    where $A=\exp{(\beta \mu)}$ is the fugacity and $a$ is the area of the system, respectively. After converting this integral into the Fermi integral function (see Appendix~\ref{App_B}), we get the expression of number density as 
    \begin{equation}
    n_g^{2D} = \frac{N}{a} = \frac{N_s}{2 \pi v_g^2} f_2\left(A\right) T^2.
    \end{equation}
    Now from the Eq.~(\ref{rohan}), the energy density for (2D) graphene system is,

    \begin{equation}
    \epsilon_g^{2D} = T_0^{0 0} =  N_s \int \frac{d^2 p}{\left(2\pi\right)^2}\left(E \right) f_0.
    \end{equation}
    After using the graphene dispersion relation and plugging the value of $f_0$, we get
    \begin{equation}
    \epsilon_g^{2D} = \frac{1}{\pi v_g^2} \int_{0}^{\infty}  \frac{E^2}{A^{-1}e^{\beta E} + 1} dE
    \end{equation}
    and after replacing this integral in terms of Fermi integral function, the final expression of the energy density is given by
    \begin{equation}
    \epsilon_g^{2D} = \frac{N_s}{\pi v_g^2} f_3\left(A\right) T^3.
    \end{equation}
    Now again from the Eq.~(\ref{rohan}), the pressure can be expressed as 
    \begin{equation}
    P_g^{2D} = T_0^{11} = N_s \int \frac{d^2 p}{\left(2\pi\right)^2}\left(\frac{E}{2} \right) f_0,
    \end{equation}
    since $\vec{p}_x \vec{v}_x \approx \frac{|\vec{p}|}{\sqrt{2}} \frac{|\vec{v}_g|}{\sqrt{2}}=\frac{E}{2}$.
    After solving this expression as similar to the energy density, we get
    \begin{equation}
    P_g^{2D} = \frac{N_s}{2 \pi v_g^2} f_3\left(A\right) T^3.
    \end{equation}
    In terms of number density, energy density, and pressure, we can write the entropy density from the Euler thermodynamic relation in 2D graphene system:
    \begin{equation}
    s = \frac{S}{a} = \frac{\epsilon + P - \mu n}{T}.
    \label{viral}
    \end{equation}
    After substituting the value of energy density ($\epsilon_g^{2D}$), pressure ($P_g^{2D}$), and number density ($ n_g^{2D}$) in Eq.~(\ref{viral}), we get
    \begin{equation}
    s_g^{2D} = \frac{N_s}{2 \pi v_g^2} T^2 \Big[3f_3\left(A\right) - \frac{\mu}{T} f_2\left(A\right) \Big].
    \label{pure}
    \end{equation}
%
%
    \subsection{Shear viscosity in two-dimensional graphene}
    Next, let us come to dissipative part of $T_D^{\mu\nu}$, where only shear stress tensor $\pi^{\mu\nu}$ will be considered for calculating shear viscosity coefficient ($\eta$). The detailed description of relativistic hydrodynamics for calculating transport coefficients can be found in Refs.~\cite{QGP1,QGP3,QGP8},
    whose graphene or unconventional version (neither relativistic nor nonrelativistic) is considered here.
    The dissipative term of energy-momentum tensor includes shear stress $\pi^{\mu\nu}$ and bulk pressure $\Pi$ as
    \be
    T^{\mu\nu}_D = \pi^{\mu\nu}+ \Pi \Delta^{\mu\nu}.
    \ee
    Here, we have assumed Landau-Lifshitz definition of flow, where $T^{\mu\nu}_D$ will be orthogonal to fluid velocity $u^\mu$, i.e. $u_\mu T^{\mu\nu}_D=0$.
    These shear stress $\pi^{\mu\nu}$ and bulk pressure $\Pi$ have proportional relations with fluid velocity gradient as
    \bea
    \pi^{\mu\nu} &=& \eta \mathcal{U}^{\mu\nu},
    \nn\\
    \Pi &=&\zeta {\cal U}_\zeta,
    \eea 
    where proportional constants are shear viscosity $\eta$ and bulk viscosity $\zeta$. Their respective fluid velocity gradients are 
    \be 
    \mathcal{U}^{\mu\nu}_\eta= D^\mu u^{\nu} +D^\nu u^{\mu} - \frac{2}{3} \Delta^{\mu\nu} \del_\rho u^\rho.
    \label{u_em1}
    \ee 
    and 
    \be 
    {\cal U}_\zeta \equiv \del_\rho u^\rho,
    \ee 
    where $\Delta^{\mu\nu}=-g^{\mu\nu}+u^\mu u^\nu$ is projection tensor normal to $u^\mu$
    and $D^\mu=\del^\mu - u^\mu u^\rho \del_\rho$ is derivative normal to $u^\mu$. They are designed
    such a way that we can get $\Delta^{\mu\nu}\equiv \delta^{ij}$ and $D^\mu\equiv \del^i$ in fluid rest frame $u^{\mu}\equiv (1, {\vec 0})$.
    Usually, greek index like $\mu\equiv(0,i)$ takes values 0 for the temporal component and $i=1,2,3$ for the spatial component for 3D system but here for 2D system, we will consider $\mu\equiv(0,i=1,2)$ because z-component $i=3$ will not be considered.
    During the transition from $\mu$-index to spatial component $i$, one can get disspative part of energy momentum tensor
    as
    \bea
    T^{ij}_D &=& \pi^{ij}+ \Pi \delta^{ij}
    \nn\\
    &=& \eta (\del^i u^j +\del^j u^i -\frac{2}{3}\delta^{ij}\nabla \cdot u)+\zeta \delta^{ij}\nabla \cdot u,
    \eea
    which ensure that diagonal part ($i=j$) is linked with bulk viscosity $\zeta$ and off-diagonal part $(i\neq j)$
    is linked with shear viscosity $\eta$. Present work will focus only on shear viscosity coefficients, so 
    we will not proceed with discussion of bulk viscosity further.

    The microscopic theory describes shear stress tensor in terms of particle velocity $v$ and momentum $p$ as,
    \be
    \pi_{\mu\nu}=N_s \int \frac{d^2 \vec{p}}{(2 \pi)^2} p_{\mu} v_{\nu} \delta f_\eta,
    \label{pij_micro}
    \ee
    where we are assuming that equilibrium distribution function $f_0$ gets a small deviation $\delta f$, which can be considered as first-order Taylor series expansion equilibrium distribution function:
    \bea 
    \delta f &\propto& \frac{\partial f_0}{\partial E}
    =\phi_\eta \frac{\partial f_0}{\partial E}
    \nn\\
    &=&\mathcal{A}^{\mu\nu} \mathcal{U}_{\mu\nu} \frac{\partial f_0}{\partial E}.
    \eea
    Considering the relation $v_{\nu}=(E/p^2) p_{\nu}$, macroscopic $\pi_{\mu\nu}=\eta \mathcal{U}_{\mu\nu}$ and microscopic Eq.~(\ref{pij_micro}) can be connected as, 
    \bea
    	\pi_{\mu\nu} &=& \eta ~\mathcal{U}_{\mu\nu}\nn\\
    	&=& N_s \int \frac{d^2 \vec{p}}{(2 \pi)^2} \left(\frac{E}{p^2}\right)p_{\mu} p_{\nu} \mathcal{A}^{\alpha\beta} \mathcal{U}_{\alpha\beta} \frac{\partial f_0}{\partial E}.
    \eea
    The four-momentum of an electron can be defined as $p^\mu=(E/v_g, \vec{p})$ in unconventional notation.
    Considering energy as a static limit of $p^\nu u_\nu$, we can write FD as,
    \be
    f_0=\frac{1}{exp{\left( \frac{p^\nu u_\nu - \mu(x)}{T(x)}\right)+1}}.
    \label{a1}
    \ee
    Here, we have to consider the local thermalization concept, where thermodynamical quantities $T(x)$, $\mu(x)$ as well as fluid velocity $u^{\mu}(x)$ are assumed to be functions of $x \equiv x^{\mu}=(x^0, x^i)$.
%
    To find the unknown coefficient $\mathcal{A}^{\alpha\beta}$, we will use Boltzmann transport equation (BTE)
    \bea
    \frac{\partial f}{\partial t}+v^\mu \frac{\partial f}{\partial x^\mu}+F^\mu \frac{\partial f}{\partial p^\mu}=\left( \frac{\partial f}{\partial t}\right)_{Col},
    \eea
    where $\Big( \frac{\partial f}{\partial t}\Big)_{Col}$ is the collision term that leads the system out of equilibrium. $F^\mu $ is represented as all external forces, and $v^\mu$ is the velocity of the fluid particles.
    Using velocity expression in terms of $E$ and $p$ for graphene, $v^\mu=(\frac{E}{p^2})p^\mu$, we get:
    \bea
    \left(\frac{E}{p^2}\right)p^\mu \partial_\mu f=\left( \frac{\partial f}{\partial t}\right)_{Col},
    \eea
    where we ignore $\frac{\partial f}{\partial t}$ and $F^\mu \frac{\partial f}{\partial p^\mu}$ as they will not contribute in shear dissipation. Using the Relaxation Time Approximation (RTA) method, the collision term can be considered as,
    \be
    \left( \frac{\partial f}{\partial t}\right)_{Col}=-\frac{\delta f}{\tau_c},
    \ee
    where $\tau_c$ is the relaxation time. Putting $f\approx f_0$ in the left-hand side (lhs) of BTE,
    \bea
    \left(\frac{E}{p^2}\right) p^\mu  \partial_\mu f_0=-\frac{\delta f}{\tau_c}.
    \label{fo}
    \eea
    Using the Eq.~(\ref{a1}), the lhs of the above Eq.~(\ref{fo}) can be expanded as, 
    \begin{align}
        \left( \frac{E}{p^2}\right) p^\mu \partial_\mu f_0 &= -f_0(1-f_0) \Big[\left(\frac{E}{p^2}\right)\frac{p^\mu p^\nu}{T} \partial_\mu u_\nu (x) \Big]\nn\\
        &=-f_0(1-f_0)  \left(\frac{E}{p^2}\right)\frac{p^\mu p^\nu}{2T} ( \partial_\mu u_\nu +\partial_\nu u_\mu)\nn\\
        &=-f_0(1-f_0) \left(\frac{E}{p^2} \right)\frac{p^\mu p^\nu}{T} \mathcal{U}_{\mu \nu},
    \end{align}
%
%
    and right hand side (rhs) of Eq.~(\ref{fo}) can be written as
    \begin{eqnarray}
    -\frac{\delta f}{\tau_c}&=& \frac{ f_0(1-f_0)}{T} \frac{1}{\tau_c} \mathcal{A}^{\mu \nu} \mathcal{U}_{\mu \nu}.
    \end{eqnarray}
%
%
    So, equating lhs and rhs of Eq.~(\ref{fo}), we get $\mathcal{A}^{\mu \nu}=(-\frac{E}{p^2} p^{\mu} p^{\nu} \tau_c)$. Transforming temporal+spatial to only spatial index, we can write the shear stress tensor as
    \bea
        \pi_{ij}&=& \eta \mathcal{U}_{ij}~,\nn\\
        &=& N_s \int \frac{d^2 \vec{p}}{(2 \pi)^2} \left( \frac{E}{p^2}\right)^2 \tau_c (p_i p_j p_k p_l) \mathcal{U}^{kl} \beta f_0(1-f_0)\nn\\
        &=& \frac{N_s}{8} \int \frac{d^2 \vec{p}}{(2 \pi)^2} E^2 \tau_c \beta f_0(1-f_0) \mathcal{U}_{ij},
    \eea
    where, we used $<p_i p_j p_k p_l>=\frac{\vec{p}^4}{8} \big(\delta_{ij} \delta_{kl}+\delta_{ik}\delta_{jl}+\delta_{il} \delta_{jk} \big)$ (see Appendix~\ref{App_D}), and we have the two Eqs;
    \begin{equation}
    \beta f_0 \left(1-f_0\right) = -\frac{\partial f_0}{\partial E}
    \label{waw1}
    \end{equation}
    and
    \begin{equation}
    -\frac{\partial f_0}{\partial E} = \beta \frac{ e^{\beta \left(E - \mu\right)}}{\left(e^{\beta \left(E - \mu\right)} + 1\right)^2} =  \frac{\partial}{\partial \mu} \left( \frac{1}{e^{\beta \left(E - \mu \right)} + 1}\right).
    \label{waw2}
    \end{equation}
 Finally, the expression for shear viscosity is,
    \begin{equation}
    \eta= \frac{N_s}{8} \int \frac{d^2 \vec{p}}{(2 \pi)^2} E^2 \tau_c \beta f_0(1-f_0).
    \label{GEOSV3D}
    \end{equation}
    After using the Eqs.~(\ref{waw1}) and (\ref{waw2}) and converting the momentum terms into energy using dispersion relation (\ref{nanu}), the Eq.~(\ref{GEOSV3D}) becomes
    \begin{equation}
    \eta = \frac{N_s }{16 \pi v_g^2} \tau_c \frac{\partial}{\partial \mu } \int_0^{\infty} \frac{E^3}{A^{-1} e^{\beta E} +1} dE.
    \end{equation}
    After solving this integration by using the identity of the  Fermi integral function, the expression of shear viscosity for 2D graphene (using subscript and superscript notation to distinguish the expressions of different systems) is
    \begin{equation}
    \eta_g^{2D} = \frac{3 N_s }{8 \pi v_g^2}\tau_c f_3\left(A\right) T^3.
    \label{gec1}
    \end{equation}
    Now, on taking the ratio of the shear viscosity (\ref{gec1}) and entropy density (\ref{pure}), we get 
    \begin{equation}
    \frac{\eta_{g}^{2D}}{ s_{g}^{2D}} = \frac{3 }{4}\tau_c \Bigg[3f_3\left(A\right) - \frac{\mu}{T} f_2\left(A\right) \Bigg]^{-1} f_3\left(A\right) T.
    \label{2dg}
    \end{equation}
    After doing a similar way of calculation, the expressions of entropy density, shear viscosity, and the ratio of shear viscosity and entropy density for a nonrelativistic electron fluid (i.e. $E=p^2/(2m)$) 2-dimensional system are given by
    \begin{align}
    & s_{NR}^{2D} = \frac{N_s m T}{2 \pi } \Bigg[2f_2\left(A\right) - \frac{\mu}{T} f_1\left(A\right) \Bigg],
    \label{pure1}\\
    & \eta_{NR}^{2D} = \frac{N_s m }{8 \pi}\tau_c f_2\left(A\right) T^2,
    \label{e_NR}\\
    & \frac{\eta_{NR}^{2D}}{ s_{NR}^{2D}} = \frac{1}{4} \tau_c \Bigg[2f_2\left(A\right) - \frac{\mu}{T} f_1\left(A\right) \Bigg]^{-1} f_2\left(A\right) T.
    \label{es_NR}
    \end{align}
    Most of the fluids or liquids (e.g., water) used in our daily life follow nonrelativistic fluid dynamics, whose fluid constituent particles obey $E=p^2/(2m)$ dispersion relation. However, for the purpose of comparing, we may assume a hypothetical 2D NR system showing fluid behavior, which may be difficult to be found in the real world. By this comparison (given details in the results section), our aim is to encourage the scientific community to use the expressions of G-case, given in Eqs.~(\ref{gec1}), (\ref{2dg}) instead NR-case, given in Eqs.~(\ref{e_NR}), (\ref{es_NR}) when they are describing eHD in graphene system.

    If we consider graphene as a 3-dimensional (3D) system, following the dispersion relation $E=pv_g$, then it may be again a hypothetical example but a good example for comparison purpose. Modifying our above calculation with replacement of $\int d^2p\rightarrow \int d^3p$ and $p_xv_x\approx \frac{pv_g}{3}=\frac{E}{3}$, we get the expressions of entropy density, shear viscosity, and their ratio as
    \begin{align}
    & s_g^{3D} = \frac{N_s T^3}{\pi^2 v_g^3} \Bigg[4f_4\left(A\right) - \frac{\mu}{T} f_3\left(A\right) \Bigg],
    \label{sg_3D}
    \\
    & \eta_{g}^{3D} = \frac{4 N_s }{5 \pi^2 v_g^3} \tau_c f_4\left(A\right) T^4,
    \label{eg_3D}
    \\
    & \frac{\eta_g^{3D}}{ s_g^{3D}} = \frac{4}{5} \tau_c \Bigg[4f_4\left(A\right) - \frac{\mu}{T} f_3\left(A\right) \Bigg]^{-1}  f_4\left(A\right) T.
    \label{esg_3D}
    \end{align}
%
    Now we have a 3-dimensional nonrelativistic (3D-NR) system of fermions. After applying the same methodology to this system, we get all the expressions of entropy density, shear viscosity, and the ratio of shear viscosity to entropy density,
    \begin{align}
    & s_{NR}^{3D} = N_s \left( \frac{m}{2 \pi }\right)^{\frac{3}{2}} T^{\frac{3}{2}} \Bigg[\frac{5}{2} f_{\frac{5}{2}}\left(A\right) - \frac{\mu}{T} f_{\frac{3}{2}}\left(A\right) \Bigg],
    \label{sNR_3D}
    \\
    & \eta_{NR}^{3D} = \frac{N_s}{4} \left( \frac{m}{2 \pi }\right)^{\frac{3}{2}} \tau_c f_{\frac{5}{2}}\left(A\right) T^{\frac{5}{2}},
    \label{eNR_3D}
    \\
     & \frac{\eta_{NR}^{3D}}{ s_{NR}^{3D}} = \frac{1}{4} \tau_c \Bigg[\frac{5}{2} f_{\frac{5}{2}}\left(A\right) - \frac{\mu}{T} f_{\frac{3}{2}}\left(A\right) \Bigg]^{-1} f_{\frac{5}{2}}\left(A\right) T.
     \label{esNR_3D}
    \end{align}
    This 3D-NR system, showing fluid behavior, can be applicable for most of the fluids or liquids (e.g. water) used in our daily life.  
    One can consider the above shear viscosity, entropy density, and their ratio for the water molecule, where water molecule obeying NR dispersion relation, $E=p^2/(2m)$ with effective mass $m$, but that will be not a good example to compare with the same expressions for eHD in graphene case. So, we can again consider a hypothetical example - 3D eHD NR case.

    We can also compare the above expressions for 2D, 3D eHD of G and NR cases with the same for ultra-relativistic (UR) case. A good example is hot QGP, where RHD can be applicable. According to the latest understanding~\cite{VR_Rev}, RHD is quite successful in describing QGP phenomenology. Again to make our comparison on equal footing, we will consider the hypothetical case of 2D, 3D-UR electron fluid.
    If the Fermi velocity of electrons in graphene $v_g$ is replaced by the factor $1$ (as $c=1$ in natural unit), then all expressions of the G-case will be converted to corresponding expressions of UR case.
%
    \section{Results}
    \label{sec: Result}
    After addressing the final expressions of $\eta$, $s$, and its ratio for different systems like 2D, 3D-NRF, GF and URF in the formalism section, here we will discuss their numerical estimations through different graphs. 
    \begin{figure}
    \centering
    \includegraphics[scale=0.3]{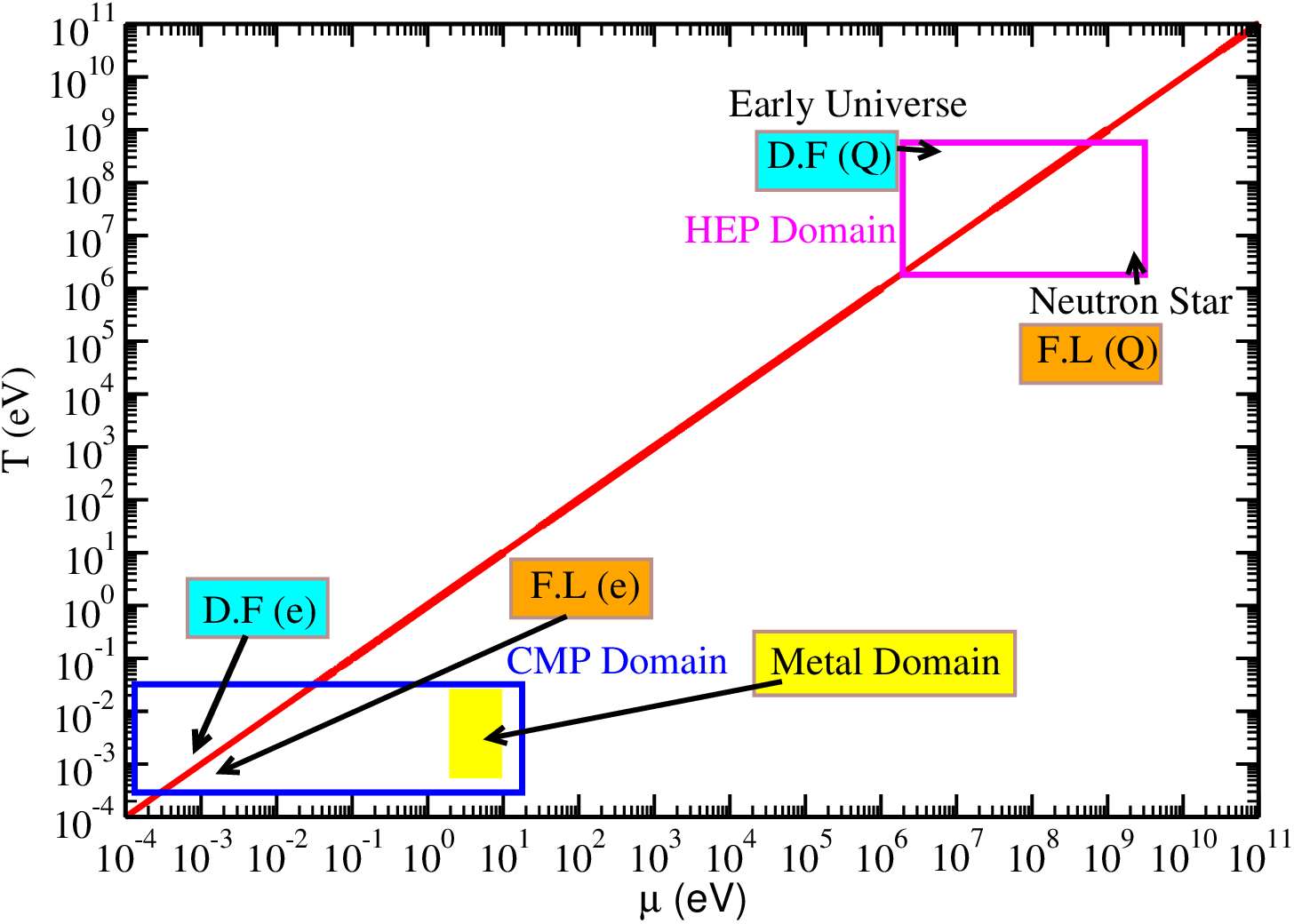}
    \caption{Location of condensed matter physics (CMP) domain and the high energy physics (HEP) domain in $T$-$\mu$ diagram.}
    \label{fig:mu_T}
    \end{figure}

    Let us first come to the entropy density results. In the early universe scenario, a hot quark-gluon plasma (QGP) state around temperature $T=400$ MeV or $T=700$ MeV and zero quark chemical potential ($\mu=0$) is expected just after a few microseconds from the big bang. Due to the very high temperature of the medium, the constituent particle average momenta become so large that we can ignore its mass term, and it can be considered a UR case. UR case is famous for photon gas or black body radiation example where internal energy density or Intensity (they have connecting relation) follows $T^4$ law, popularly known as Stefan-Boltzmann (SB) law. QGP thermodynamics at high temperatures reaches that SB limits. Eq.~(\ref{sg_3D}) can be converted to UR case by replacing $v_g=c=1$ and by putting $\mu=0$, we can get SB limit expression for 3D case,
    \be 
    s_{SB} = \frac{N_s T^3}{\pi^2 } \Bigg[4\times \Big\{\frac{7}{8}\zeta_4\Big\} \Bigg],
    \label{sSB_3D}
    \ee 
    where the Fermi integral function can be converted into Riemann zeta function $f_4=\frac{7}{8}\zeta_4$ for $\mu=0$ condition, following the general relation $f_n=\Big(1-\frac{1}{2^{n-1}}\Big)\zeta_n$. For the QGP case, quark degeneracy factors will have to be put into the $N_s$, and gluon contribution must be added separately. Since two-flavor quark has a degeneracy factor of 24 and gluon has a degeneracy factor of 16, so massless QGP entropy density or SB limits of QGP will be 
    \be 
    s^{QGP}_{SB} = \frac{24 T^3}{\pi^2 } \Big[4\times \Big\{\frac{7}{8}\zeta_4\Big\} \Big]+\frac{16 T^3}{\pi^2 } \Big[4\times \zeta_4 \Big].
    \label{s_QGP}
    \ee 
    When we plan to compare graphene entropy density with this SB limit, we have to understand that the temperature range (few hundred mega electron volt (MeV), which is equivalent to $10^{12}~^0$K ) of QGP is too much larger than temperature range (1-23 milli-electron volt (meV), which is equivalent to 15-300 $^0$K ) of graphene system. Fig.~(\ref{fig:mu_T}) has addressed nicely about these two domains. It is basically $T$ vs. $\mu$ plots in log scale to cover a broad band of $T$ and $\mu$ range. We marked the condensed matter physics (CMP) domain, covering  $T\approx 1-23$ meV and $\mu\approx 0-10$ eV. We know that metal Fermi energy remains within the range $\mu=2-10$ eV, which is marked as yellow. Unlike metal, graphene system Fermi energy can be changed via doping methods, and its $\mu/T\ll 1$ and $\mu/T\gg 1$ domains are called Dirac fluid (DF) or Dirac liquid (DL) and Fermi liquid (FL) domains, respectively, marked by arrows in Fig.~(\ref{fig:mu_T}). Similar to DF and FL domains for electrons, we may call early universe QGP as DF domain of quark and quark matter, expected in the core of neutron star as FL domain of quark. A rectangular domain within $T=1-400$ MeV and $\mu=0-1000$ MeV is marked as high energy physics (HEP) domain for quark. Reader can easily visualize the gap between CMP and HEP domains.
    \begin{figure*}
    \centering
    \includegraphics[scale=0.3]{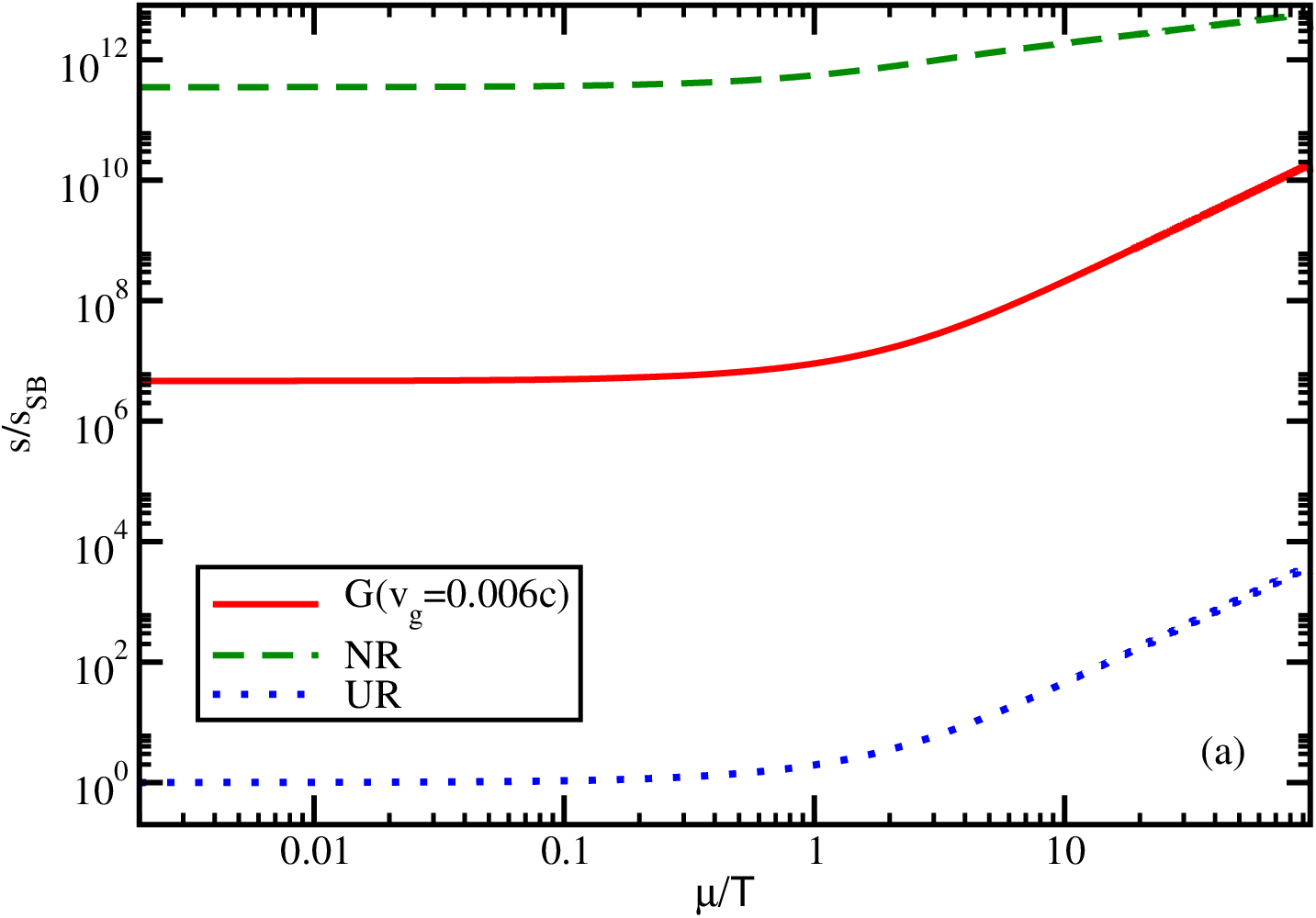}
    \includegraphics[scale=0.3]{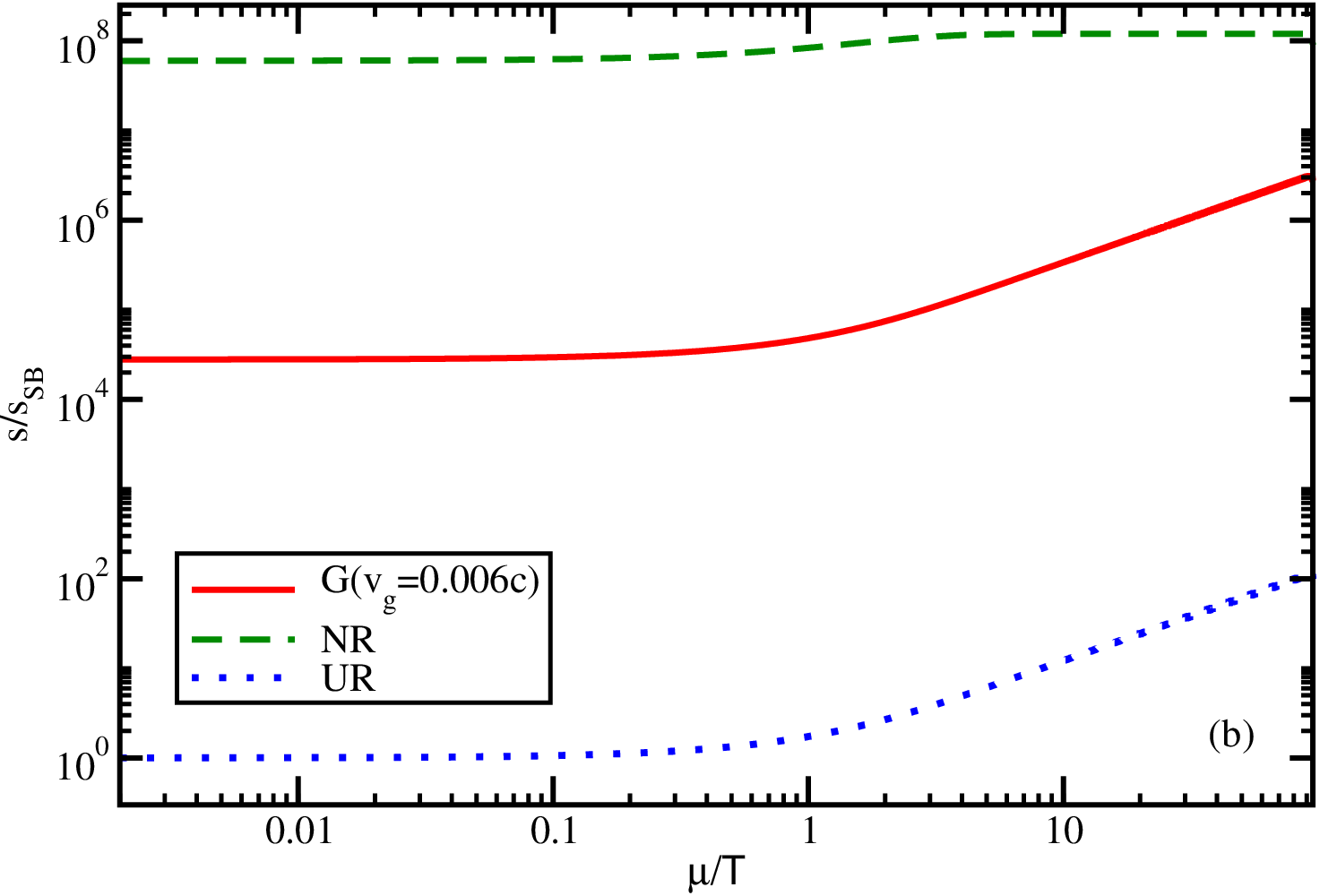}
    \caption{The ratio of entropy density in different domains to $s_{SB}$ with $\mu/T$ (a) in the 3D case and (b) in the 2D case respectively}
    \label{fig:s}
    \end{figure*}
    After realizing the scale gap in $T$-$\mu$ plane between URF of quark and GF of electrons, one should understand that we must consider a hypothetical electron URF to make an equal footing comparison. Within the temperature ($T=0-0.023$ eV) and chemical potential ($\mu=0-10$ eV) range, entropy density of URF,
    \be 
    s_{UR}^{3D} = \frac{N_s T^3}{\pi^2} \Bigg[4f_4\left(A\right) - \frac{\mu}{T} f_3\left(A\right) \Bigg],
    \ee 
    has to be plotted with a normalization by $s_{SB}$, given in Eq.~(\ref{sSB_3D}). This normalized estimation is sketched by the blue dotted line in the left panel of Fig.~(\ref{fig:s}), which shows that $s_{UR}^{3D}\approx s_{SB}$ in the domain $\mu/T\ll 1$, as expected. Interestingly, we noticed that the main $\mu/T$ dependence in entropy density is coming beyond the $\mu/T=1$. Reader can understand that the terms with Fermi integral function are the main source of $\mu/T$ dependence. Next, using Eq.~(\ref{sg_3D}), the graphene entropy density for the Fermi velocity $v_g=0.006$ is plotted (red solid). From Ref.~\cite{vg_Nature}, we can get knowledge about a broad range of Fermi velocity $v_g=1-3\times 10^6$ m/s or $v_g=0.003-0.01$ (in natural unit) in graphene system. As charge career density or $\mu$ decreases, $v_g$ will increase and approach towards Dirac fluid (DF) or strongly coupled electron-electron domain. We have considered in-between constant values $v_g=0.006$. We can understand that the $\mu/T$ dependence of entropy density for URF and GF are the same but $GF\gg URF$ due to the $1/v_g^3\approx 5\times 10^6$ term. Next, we use Eq.~(\ref{sNR_3D}) to draw the entropy density of NRF to plot (green dash line) in the left panel of Fig.~(\ref{fig:s}). Reader can understand its different trend of $\mu/T$ dependence for NRF is because of the term $\Big[\frac{5}{2} f_{\frac{5}{2}}\left(A\right) - \frac{\mu}{T} f_{\frac{3}{2}}\left(A\right) \Big]$.

    A similar trend we can notice for 2D case with similar ranking URF $\ll$ GF $\ll$ NRF. Only for the transition from 3D to 2D, their orders of magnitude are shifted toward lower values. 

\begin{figure*}
    \centering
    \includegraphics[scale=0.3]{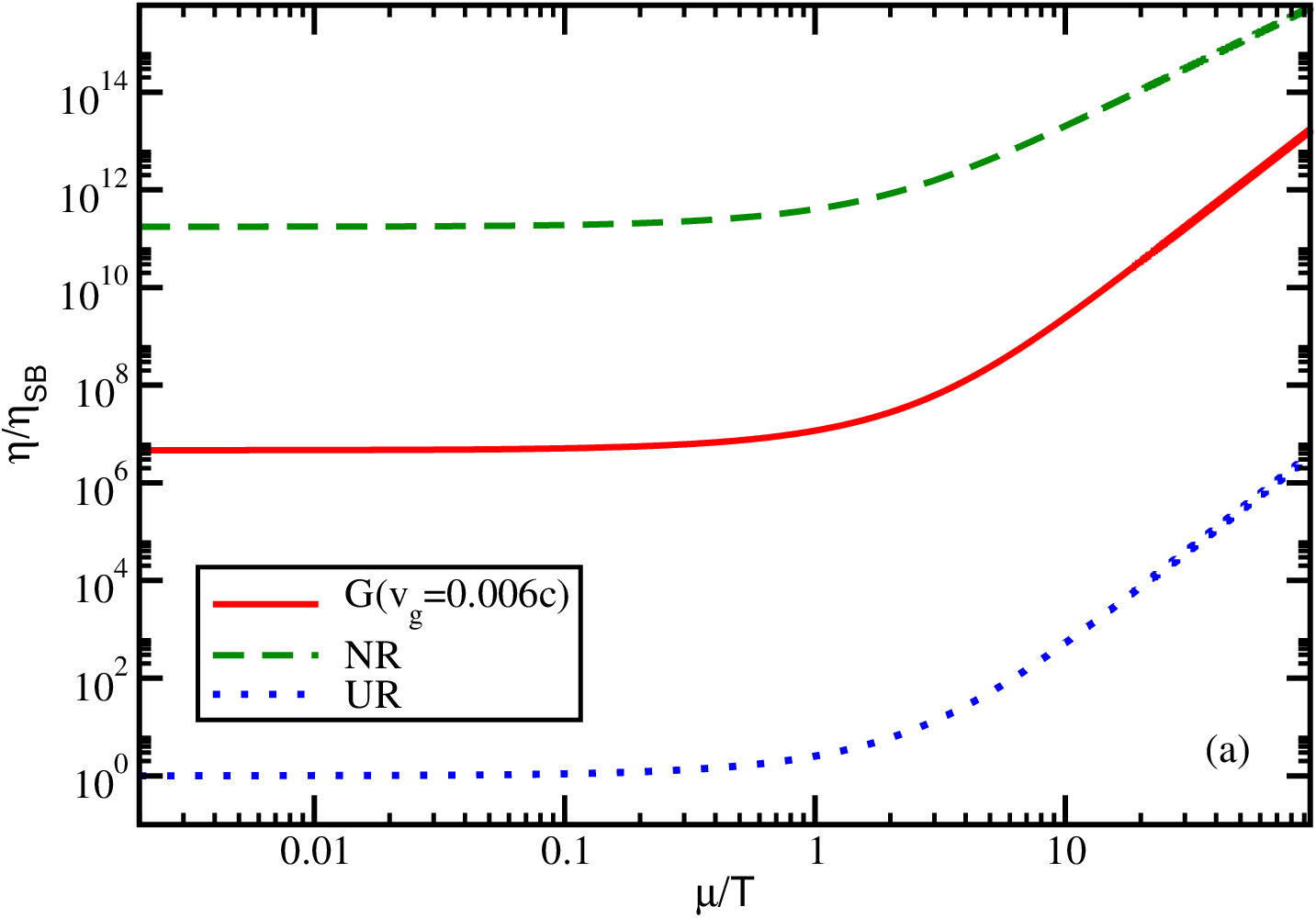}
    \includegraphics[scale=0.3]{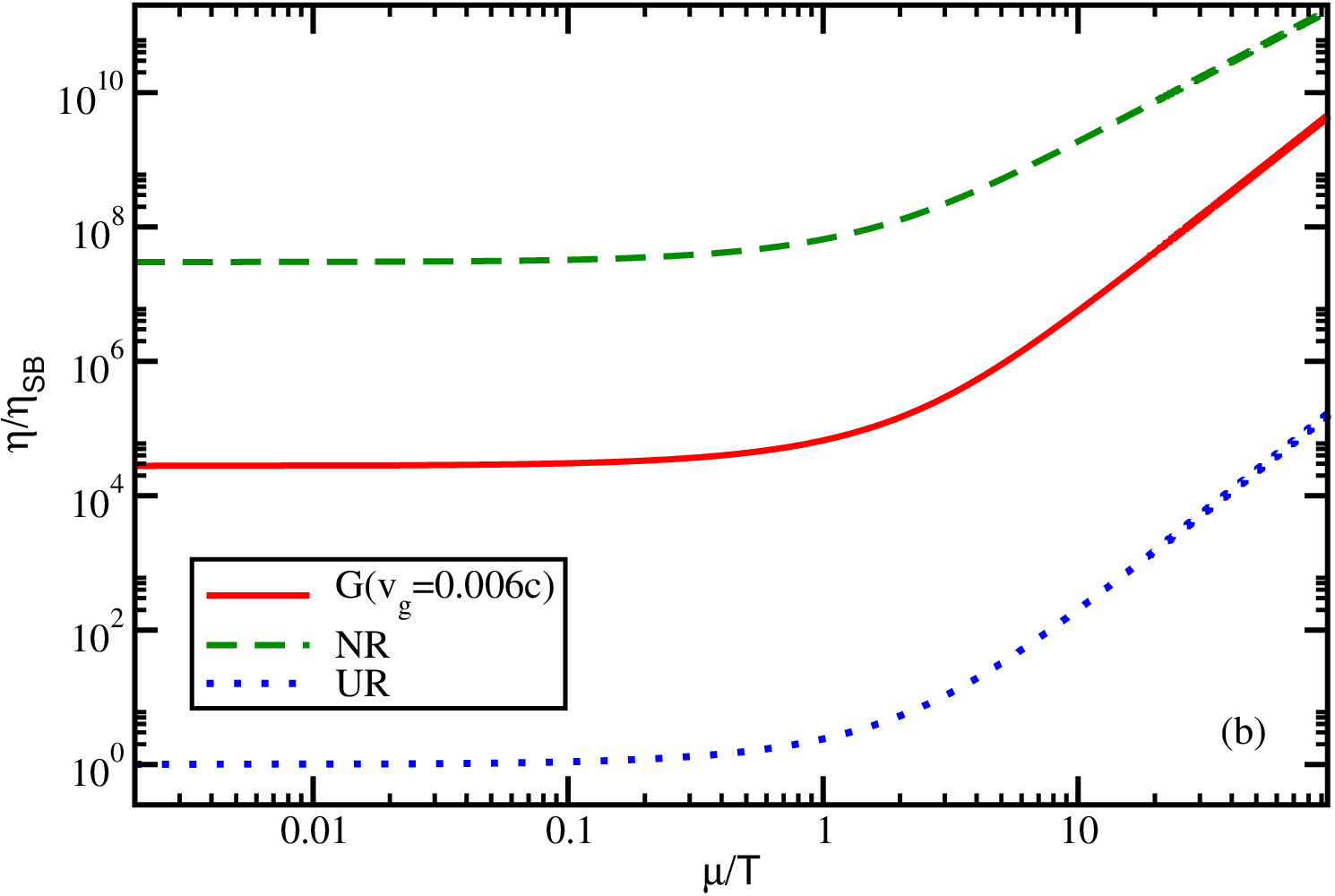}
    \caption{The ratio of shear viscosity of electron flow in different domains to $\eta_{SB}$ with $\mu/T$ (a) in the 3D case and (b) in the 2D case respectively}
    \label{fig:e}
    \end{figure*}

    Next, let us come to the shear viscosity results. Here also, we can expect SB limit type simple expression for UR case at $\mu=0$:
    \be 
    \eta_{SB} = \frac{4 N_s }{5 \pi^2} \tau_c \frac{7}{8}\zeta_4 T^4,
    \label{eSB_3D}
    \ee 
    from the general $\eta(T,\mu)$ expression for URF:
    \be 
    \eta_{UR}^{3D} = \frac{4 N_s }{5 \pi^2} \tau_c f_4\left(A\right) T^4,
   \label{eUR_3D}
    \ee 
    by putting $v_g=c=1$ in Eq.~(\ref{eg_3D}).
    For massless QGP at $\mu=0$ case, by replacing degeneracy factors of quarks and gluons in $N_s$, we get 
    \be 
    \eta^{QGP}_{SB} = 24\Big[\frac{4}{5\pi^2 } \tau_c . \Big\{\frac{7}{8}\zeta_4\Big\}T^4 \Big]+16\Big[\frac{ 4}{5\pi^2 } \tau_c . \zeta_4 T^4 \Big].
    \ee
    Again, this QGP is a realistic example of URF but for comparison, we have to consider electron URF. When we plan to compare the shear viscosity of URF, GF, and NRF, then we should use Eqs.~(\ref{eUR_3D}), (\ref{eg_3D}), (\ref{eNR_3D}) and for SB limit, we will use Eq.~(\ref{eSB_3D}). Similar to normalized entropy density by its SB limit in Fig.~(\ref{fig:s}), we have plotted normalized shear viscosity by its SB limit in Fig.~(\ref{fig:e}), where 3D and 2D estimations are plotted in left and right panels respectively. Shear viscosity expression carries two kinds of information. One is relaxation time $\tau_c$, and another is the remaining thermodynamic phase-space part as a function of $T$ and $\mu$. During the normalization, $\tau_c$ information is canceled, and we can only see their thermodynamic phase-space part of shear viscosity. Interestingly, it follows a similar trend to other thermodynamical quantities like entropy density - which shows two types of $\mu/T$ dependence in the domain $\mu/T\ll 1$ and $\mu/T\gg 1$, which are commonly assigned as DF and FL.

%
    Now, let us come to the shear viscosity to entropy density ratio $\eta/s$, which is a more important quantity than only $\eta$ to measure the fluidity of the system. In the D.F domain, the extreme situation (mathematically) is $\mu \rightarrow 0$. In this limit, $\eta$ and $s$ carry quite similar terms, so when we take their ratio, we will get very simplified expressions:
    \bea 
    \frac{\eta}{s} &=& \frac{\tau_c T}{5} ~{\rm for ~3D ~URF/GF},\\
    \frac{\eta}{s} &=& \frac{\tau_c T}{10} ~{\rm for ~3D ~NRF},\\
    \frac{\eta}{s} &=& \frac{\tau_c T}{4} ~{\rm for ~2D ~URF/GF},\\
    \frac{\eta}{s} &=& \frac{\tau_c T}{8} ~{\rm for ~2D ~NRF}.
    \eea
    From the String theory-based calculation~\cite{25Kovtun:2004de}, it was conjectured that $\eta/s$ has a lower bound, well known as KSS bound, which gives an inequality $\frac{\eta}{s} \geq \frac{\hbar}{k_B} \frac{1}{4 \pi}=\frac{1}{4 \pi} $ (in natural unit). Though classically one may expect $\tau_c \rightarrow 0 \Rightarrow \frac{\eta}{s} \rightarrow 0$, but quantum mechanically, relaxation or scattering time $\tau_c$ or mean free path $\lambda_c \approx v \tau_c$ can not be lower than de-Brogile range of time or wavelength scale. This simple quantum mechanical concept also suggests a lower bound of $\frac{\eta}{s}$, sometimes called a quantum lower bound. By imposing this bound $\frac{\eta}{s}= \frac{1}{4 \pi}$, we can get a rough expression lower bound of $\tau_c$ as
    \bea 
    \tau_c &=& \frac{5}{4 \pi T} ~{\rm for ~3D ~URF/GF},\\
    \tau_c &=& \frac{10}{4 \pi T} ~{\rm for ~3D ~NRF},\\
    \tau_c &=& \frac{4}{4 \pi T} ~{\rm for ~2D ~URF/GF},\\
    \tau_c &=& \frac{8}{4 \pi T} ~{\rm for ~2D ~NRF}.
    \label{tc_KSS}
    \eea
    \begin{figure*} 
    \centering
    \includegraphics[scale=0.3]{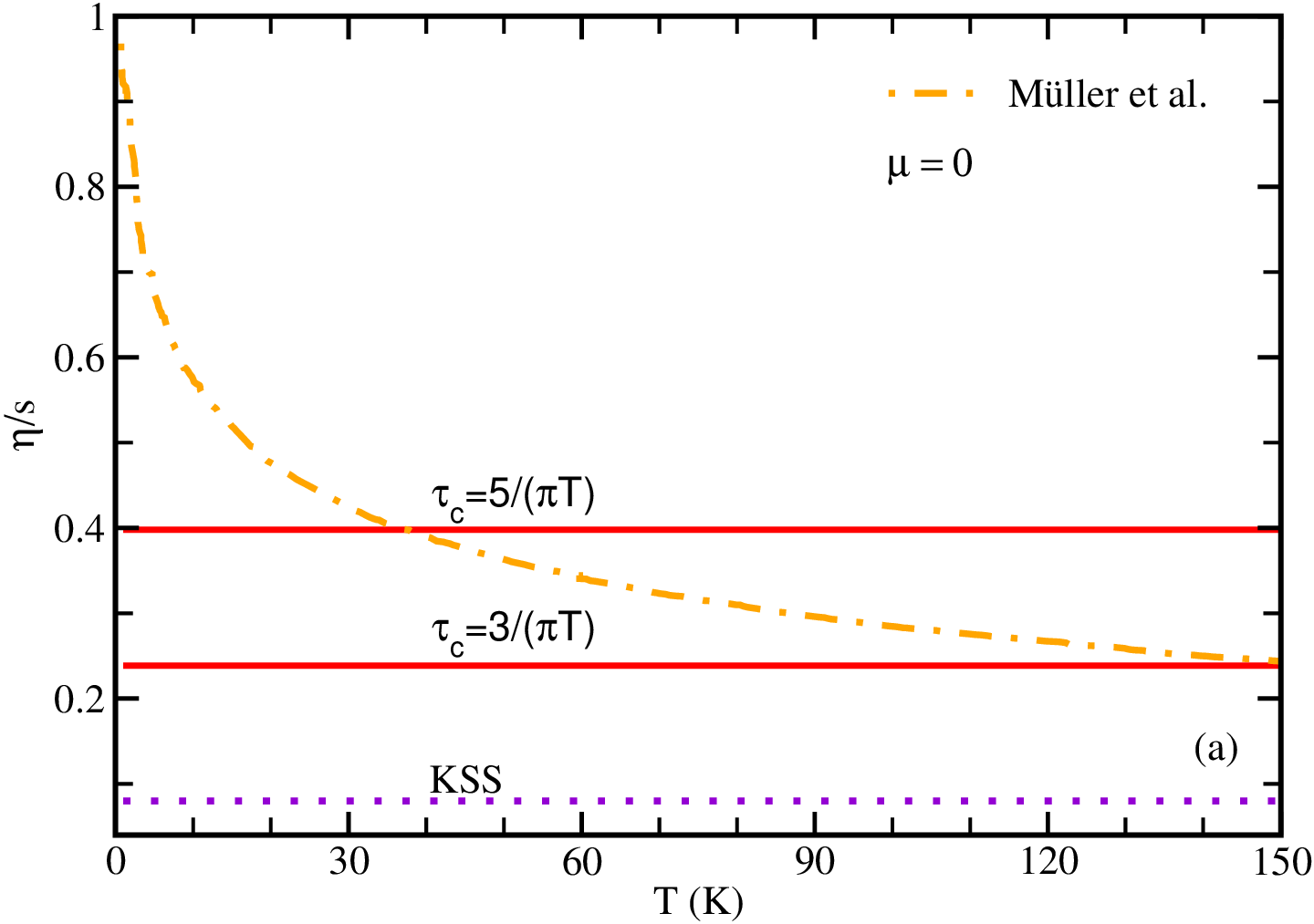}
    \includegraphics[scale=0.3]{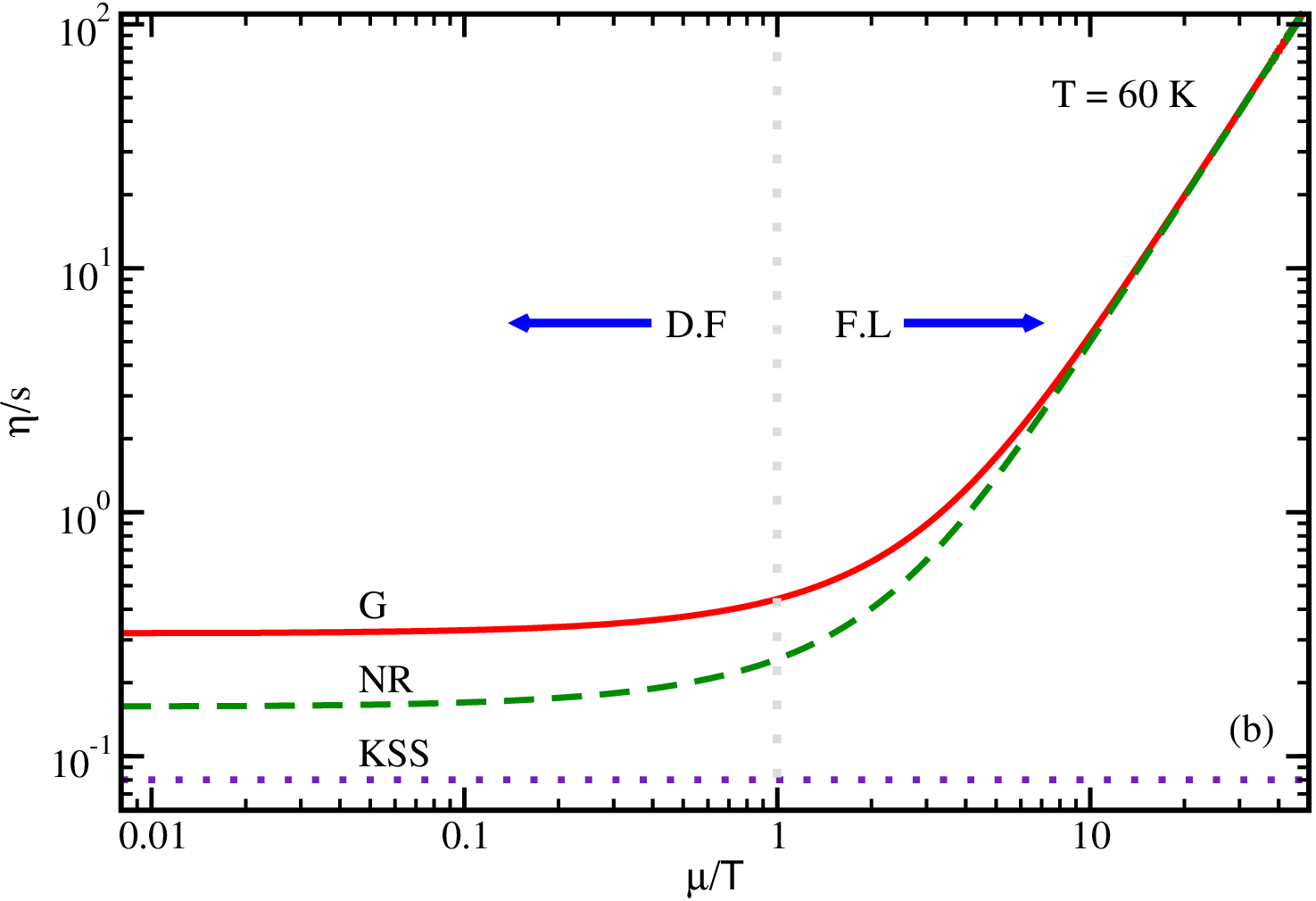}
    \caption{Shear viscosity to entropy density ratio; (a) with $T$ for $\mu=0$ (Undoped graphene) and (b) with $\mu/T$}
    \label{fig:es}
    \end{figure*}
    This KSS bound conjecture~\cite{25Kovtun:2004de} makes the scientific community curious to find such fluid whose $\eta/s$ is close to this bound. In other words, if we write $\eta/s=n/(4\pi)$, where $n\geq 1$, then fluid with $n=1-5$ may be considered as those special fluids and may be called nearly or close to perfect fluid. Empirically, QGP is the evidence of such perfect fluid ($n\approx 1-2$) in the relativistic domain, while close to perfect fluid ($n\approx 5$) example for NR case is cold atom systems~\cite{27Sch_fer_2009}. According to Eqs.~(\ref{tc_KSS}), we can expect gross values of relaxation time for QGP and cold atom systems as $\tau_c\approx\frac{5}{4\pi T}$-$\frac{10}{4\pi T}$ and $\tau_c\approx\frac{50}{4\pi T}$ respectively. 
    Similarly, according to the theoretical prediction from Ref.~\cite{24PhysRevLett.103.025301}, GF may also belong to this close-to-perfect fluid category. So far, to the best of our knowledge, no experimental measurement of $\eta/s$ vs. $T$ plot is available, so the theoretical plot of $\eta/s$ vs. $T$ in Ref.~\cite{24PhysRevLett.103.025301} is considered as our reference to guess or tune order of magnitude for $\tau_c\approx\frac{n}{\pi T}$. 
    In the left panel of Fig.~(\ref{fig:es}), we can get guidance that $\tau_c\approx\frac{n}{\pi T}$ within $n=3$-$5$ can cover the order of magnitude of $\eta/s$ in the temperature range $T=35$-$150^\circ$K, predicted by Müller {\it et al}.~~\cite{24PhysRevLett.103.025301}. By considering an average value $\tau_c\approx\frac{4}{\pi T}$, we have plotted $\eta/s$ of 2D GF or URF (red solid line) and 2D-NRF (green dash line) against $\mu/T$-axis in the right panel of Fig.~(\ref{fig:es}).
    We notice that $\eta/s$ in DF domain becomes lower than in the FL domain, mainly because of the thermodynamical phase-space part of $\eta/s$. In terms of Fermi integral function, this part for GF can be identified as $\Big[3f_3\left(A\right) - \frac{\mu}{T} f_2\left(A\right) \Big]^{-1} f_3\left(A\right)$ from Eq.~(\ref{2dg}).
    We have put the NRF case for reference, but 2D-NRF for electrons may be possible in a hypothetical situation. So present study indicates that dropping $\eta/s$ values and saturating towards constant values may be found during the transition from FL to DF domains in the graphene system.
    Although we have a limitation in that we have considered $\tau_c\propto 1/T$, which may be changed in the actual microscopic calculation of $\tau_c$, and so the trend of $\eta/s$ may also be changed. It demands more theoretical studies on these $\eta/s$ estimations as well as the explicit measurement of this quantity from the experimental side.

    The inversely proportional relation ($\tau_c=\frac{\alpha^2}{T}$, where $\alpha$ is effective fine structure constant) between $\tau_c$ 
    and $T$ can also be noticed in the work of Müller {\it et al}.~\cite{24PhysRevLett.103.025301}, which reflects that the KSS bound inspired expression 
    of relaxation time $\tau_c\propto\frac{1}{T}$ is also in good agreement with microscopic calculations~\cite{24PhysRevLett.103.025301}.
    A proportional relation $\tau_{c}\propto T$ is also estimated from the AdS/CMT based calculations~\cite{PhysRevD.88.086003,Andrade2014-fk,Davison2015-ow}, 
    whose $m$ of expression $\tau_c=1/\Gamma = \frac{4\pi T}{m^2}$ is basically an effective mass parameter for some of the metric 
    fluctuations in the gravitational theory. However, one may get $\tau_c=1/\Gamma = \frac{4\pi}{T} \propto \frac{1}{T}$ dependence in
    the condition $m \propto T$. In searching for the exact $T$-dependence of relaxation time $\tau_c(T)$ of electrons for graphene systems, more
    microscopic calculations in different directional (may be model dependent) attempts may be needed for enriching the understanding of this field.

    \section{Summary}
    \label{sec: Sum}
    We can summarize our investigation in the following steps. First, we introduce a brief macroscopic description of electron fluid in graphene, then we focus on its microscopic description. Our dealing quantity is considered as energy-momentum tensor, whose ideal part represents energy density and pressure in the static limit picture of fluid dynamics. Using those thermodynamical quantities, our destination from the ideal part of the energy-momentum tensor becomes entropy density, which will be used to be normalized with shear viscosity. From the dissipative part of the energy-momentum tensor, shear viscosity coefficients of electron fluid are calculated based on the kinetic theory approach with relaxation time approximation. Temperature and chemical potential-dependent general expressions of shear viscosity, entropy density, and the shear viscosity to entropy density ratio has been calculated and plotted for different cases of electron fluid like nonrelativistic, graphene and ultra-relativistic cases. For completeness of comparison, we considered both 3D and 2D systems. Analyzing the results of different cases, we get a comparative understanding and conclusions, which are addressed briefly in bullet points:
    \begin{enumerate}
    \item The $\mu/T$ dependence of shear viscosity $\eta$ as well as entropy density $s$ for URF and GF are exactly similar but a little different from the NRF.
    \item We notice a huge difference among URF, GF, and NRF in terms of the order of magnitude of $\eta$ and $s$ with ranking URF $\ll$ GF $\ll$ NRF.
    \item During transiting from 3D to 2D, order of magnitude of $\eta$ and $s$ shift towards lower values.
    \item When we go from Fermi Liquid ($\mu/T\gg 1$) to Dirac Liquid ($\mu/T\ll 1$) domain, values of $\eta$,
        $s$ and $\eta/s$ ratio decrease towards a saturated values.
    \item Interesting ranking for $\eta/s$ becomes URF $=$ GF $\ge$ NRF.
    \end{enumerate}
    
    The present comparative study on the microscopic calculation of shear viscosity may be considered as a good documentation of master formulas of different cases from 3D-URF, GF, NRF to 2D-URF, GF and NRF. In the future, it may be useful for actual graphene system estimation, where one should go with some first principle or model-dependent calculation of relaxation time. Also, one should deal with electron-hole plasma with appropriate degeneracy factor in the Dirac Fluid domain for the actual graphene system but the present work sticks with electron description only due to observing the estimations for its  different dispersion relations. Our immediate future plan is to concentrate on the actual graphene phenomenology on the viscous aspects. 

    \begin{acknowledgments}
    This work was partly (CWA and TZW) supported by the Doctoral Fellowship in India (DIA) program of the Ministry of Education, Government of India. The authors thank the other members of eHD club - Sesha P. Vempati, Ashutosh Dwibedi, Narayan Prasad, Bharat Kukkar, and Subhalaxmi Nayak.
    \end{acknowledgments} 
%
    \appendix
    \section{Density of States}
    \label{App_A}
    The density of states is nothing but the total number of energy states per unit energy interval. If the total number of energy states in energy range $E$ to $E + dE$ are $D\left(E\right)dE$, then the density of states will be 
    \begin{equation}
    g\left(E\right) = \frac{D\left(E\right)dE}{dE}~.
    \end{equation}
    Now, here are some expressions of the number of energy states for different-different cases expressed as
    \hfill \break
\\
    {\it Case:1.} for 3D-Graphene,
    \begin{equation}
    D\left(E\right)dE = N_s \frac{4\pi V}{h^3 v_g^3}E^2 dE,
    \end{equation}
    {\it Case:2.} for 3D nonelativistic,
    \begin{equation}
     D\left(E\right)dE = N_s 2\pi V \left(\frac{2m}{h^2}\right)^{\frac{3}{2}}\sqrt{E} dE,
    \end{equation}
    {\it Case:3.} for 2D Graphene,
    \begin{equation}
    D\left(E\right)dE = N_s \frac{2\pi a}{h^2 v_g^2} E \, dE,
    \end{equation}
    {\it Case:4.} for 2D Nonrelativistic,
    \begin{equation}
     D\left(E\right)dE = N_s \frac{2\pi a}{h^2} m\, dE.
    \end{equation}
    In the above expressions, $V$ represents the volume in position space.

    \section{Fermi-Dirac Integral}
    \label{App_B}
    We have the integral form as
    \begin{align}
        f_\nu (A)=\frac{1}{\Gamma (\nu)}\int_0^\infty \frac{x^{\nu-1}}{A^{-1} e^x+1} dx,
        \label{mom}
    \end{align}
    where $f_\nu (A)$ is known as the Fermi-Dirac integral and $x= \beta E$. And the expression in terms of energy can be written in terms of $x$ as
    \begin{align}
        \int_{0}^{\infty}  \frac{E^{\nu -1}}{A^{-1}e^{\beta E} + 1} dE &= \frac{1}{\beta^\nu} \int_0^\infty \frac{x^{\nu-1}}{A^{-1} e^x+1} dx\nn\\
        &= \frac{1}{\beta^\nu} \Gamma (\nu)  f_\nu (A).
    \end{align}
%
%
%
    \section{Average Angular Integral in 2D}
    \label{App_D}
    We have the integral form as
    \begin{align}
    \int p_i p_j p_k p_l d^2p = p dp \int  p_i p_j p_k p_l d\theta.
    \end{align}
    Since
    $$\Vec{p} = p \hat{n},$$
    where $$ \hat{n} = \cos{\theta}\,\hat{i} + \sin{\theta}  \hat{j}, $$
    and
    $$ p_i = \Vec{p}.\hat{e_i} = p \left(\hat{n}.\hat{e_i}\right) = pn_i, $$
    the integral becomes
    \begin{align}
     \int p_i p_j p_k p_l d^2p = p dp, p^4 \int  n_i n_j n_k n_l \,d\theta.
    \end{align}
    Now, we have to calculate
    $$ \int  n_i n_j n_k n_l \,d\theta = ? $$
    {\it Case:1.} 
    The above integral becomes
    \begin{align}
        \int n_1^2 n_2^2\, d\theta &= \int_0^{2 \pi} \cos^2{\theta} \sin^2{\theta} \, d\theta\\
        & = 4  \int_0^{ \frac{\pi}{2}} \cos^2{\theta} \sin^2{\theta} \, d\theta.
    \end{align}
    Now, using the Beta function, we know that
    \begin{align}
        & B\left( u, v\right) = 2 \int_0^{\frac{\pi}{2}} \left(\cos{\theta}\right)^{2u-1}\left( \sin{\theta}\right)^{2v-1} \, d\theta\\
        \Rightarrow & B\left( u, v\right) = \frac{\Gamma u\, \Gamma v}{\Gamma\left(u+v\right)}.
    \end{align}
    Applying this, we get
    \begin{equation}
        \int n_1^2 n_2^2\, d\theta = \frac{2 \pi}{8}.
    \end{equation}
    {\it Case:2.}
    \begin{align}
        \int n_1^3 n_2\, d\theta &= \int n_1 n_2^3\, d\theta\\
        & = \int_0^{2 \pi} \cos^3{\theta} \sin{\theta} \, d\theta = 0.
    \end{align}
    {\it Case:3.}\\
    \begin{align}
        \int n_1^4\, d\theta &=  \int n_2^4\, d\theta \\
        & = \int_0^{2 \pi} \sin^4{\theta} \, d\theta\\ 
        & = 2 \, B\left( \frac{5}{2}, \frac{1}{2}\right) = \frac{3 \pi}{4}.
    \end{align}
    Now, the above integral can be written as
    \begin{align}
        \int n_1^4\, d\theta &= \int n_2^4\, d\theta = \frac{2 \pi}{8}\times 3,
    \end{align}
    Now, the integral can be expressed as
    \begin{align}
        \int n_i n_j n_k n_l \,d\theta = \frac{2\pi }{8} \left( \delta_{ij}\delta_{kl} + \delta_{ik}\delta_{jl} + \delta_{il}\delta_{jk}\right)
    \end{align}
    and
    \begin{align}
        \int p_i p_j p_k p_l \, d^2p = 2\pi \, p\,dp \frac{p^4}{8}\left( \delta_{ij}\delta_{kl} + \delta_{ik}\delta_{jl} + \delta_{il}\delta_{jk}\right).
    \end{align}
    Now, the average final expression will be
    \begin{equation}
        <p_i p_j p_k p_l> = \frac{p^4}{8}\left( \delta_{ij}\delta_{kl} + \delta_{ik}\delta_{jl} + \delta_{il}\delta_{jk}\right).
    \end{equation}
    \bibliography{new_ref}
    \end{document}